\documentclass[aps,floats,amsmath,amssymb,twocolumn]{revtex4}
\usepackage{calc}
\usepackage{psfrag}
\usepackage{graphicx}
\usepackage{color}

\begin{document}

\begin{abstract}
The Berry curvature in Chern insulators  appears to be a non-gauge-invariant
quantity and does not immediately  allow local length characterization. However, in two examples of 2- and 3-band models that we discuss, we find high-symmetry points in the Brillouin zone that have Berry curvature invariant under diagonal gauge transformations, and may serve as expansion points of geometrical description. On the basis of the geometrical description, in the case of Dirac based 2-band Chern insulators like Haldane model we conclude that the characteristic length based on the value of the Berry curvature at the expansion point plays the role of the magnetic length  in the expression for the Hall viscosity. In the case of 2-band models the characteristic ``cyclotron" spin is equal to 1/2, while
in the 3-band kagome case this spin is likely non-quantized and non-universal.
\end{abstract}
\title{Effective description of Chern insulators}
\author{E. Dobard{\v{z}}i\'c$^1$}
\author{M. Dimitrijevi\'c$^1$}
\author{M.V. Milovanovi\'c$^2$}
\affiliation{$^1$ Faculty of Physics, University of Belgrade, 11001 Belgrade, Serbia\\
$^2$ Scientific Computing Laboratory, Institute of Physics
Belgrade, University of Belgrade, Pregrevica 118, 11 080 Belgrade,
Serbia}

\maketitle

\section{Introduction}
Chern insulators (CIs)~\cite{Haldane_parity} exhibit integer quantum Hall effect (IQHE) conductance quantization in the absence of the
magnetic field due to non-trivial filled band structure with non-zero topological Chern number. In the ordinary (continuum - not on a lattice) IQHE the uniform external magnetic field defines a characteristic length, i.e. the magnetic length that describes classically speaking ``the size of the particle orbit in the magnetic field", or better characteristic volume per particle. On the other hand, in model CIs, i.e. those based on quadratic Hamiltonians, there is no obvious way to define such a length. Although one is inclined to consider CI physics as a single particle problem in a varying magnetic field (in inverse space), which we identify with the local Berry curvature, this point of view may be questioned since the Berry curvature appears to be a gauge non-invariant quantity (as we will show below). Interactions, if relevant, may fix gauge through spontaneous symmetry breaking, i.e. when fermionic bilinears acquire non-zero values. Therefore it is not clear if it is always possible to define a local, physical characteristic length in  CIs.

On the other hand, the quantization of the particle volume in the QHE problem inspired the geometric approach to fractional QHE (FQHE)~\cite{Haldane_geo_prl}. The dynamical metric degree of freedom is constrained by the demand that the
metric is uni-modular (this signifies the quantization of the volume, i.e. commensuration of flux and particles). Low-lying collective modes are identified with the oscillations in the shape of the characteristic volume, i.e. changes in the metric, and the measure of the density variations is given by a local curvature. Thus it is natural to ask whether this view of QHE phenomena may be extended to the domain of (fractional)CIs given that  there is no obvious gauge-invariant local length characterization.

In the ordinary QHE there is a quantized response of the system where the magnetic length enters together with the characteristic spin of the QHE state~\cite{Avron,rr}. This quantity is Hall viscosity and it was explored in the context of Dirac based CIs in Refs.~\cite{Hughes1,Hughes2}. A non-universal (dependent on the model parameters) characteristic length  that takes place of the magnetic length in the expression for the IQHE system was identified.
This was achieved considering the system in the presence of a non--trivial geometric background, and analyzing its low-energy, long-wavelength response near Dirac points.
Inspired by the approach of Ref.~\cite{Maci}, Ref.~\cite{You} discussed the case of CIs based on quadratically dispersing free 2-band CI Hamiltonian in the presence of nematic ordering. The Hall viscosity was discussed, a physical (non-cut-off dependent)
quantity was not given, but one  expects that this quantity exists  with a characteristic length.
In other words in principle from the physical response we expect to recover a characteristic length and a global, i.e. large scale (long wavelength) characterization.

In this paper we will not discuss the response to geometrical perturbations. Instead we will try to understand
whether the internal degrees of freedom can be described in geometrical terms.
If this internal ``geometrization" exists its characteristic length should coincide with the characteristic length for the response to the geometrical perturbations. We will discuss specific 2-band and 3-band CI and ask whether a geometric description of a CI ground state  in which the characteristic length  plays the role of magnetic length in the expression for the Hall viscosity is possible. We find that  in some regions of parameter space
a global, geometric characterization with a characteristic length exists.

The paper is organized as follows. In Section~\ref{Berry curvature and gauge transformations} the meaning and the form of the gauge transformations in the
context of the Bloch problem will be explained.  In Section~\ref{Dirac based 2 band CIs and the geometric description of the single particle problem} the basic  elements of the geometry of the Dirac based 2 band CIs are discussed with the emphasis on the gauge-invariant Berry curvature at Dirac point, and its special relationship with the Fubini-Study (FS) metric. 
After a short overview  of the geometric description of fractional QHE in Section~\ref{The geometric description of (F)QHE - overview}, in Section~\ref{The geometric description of the ground state of interacting Dirac based CIs} the geometric description of the non-interacting CI problem with a zero-flux equation is introduced. The geometric description of the interacting CI problem was given in terms of a Lagrangian that mixes ``cyclotron" and ``guiding center" degrees of freedom. The form of the Lagrangian was later used in Section~\ref{Discussion} for a diagnostics of the characteristic length in agreement with previous  studies on the response of the system. Section~\ref{Comparison with other 2 band models} discusses 2-band models in general, and especially the demand on the expansion point to be the point of the smallest energy gap. In Section~\ref{3 band kagome model} the geometry of the 3-band kagome model is discussed, i.e.
the existence of the point with gauge invariant Berry curvature and the curvature special relationship with FS metric, which enable the geometric description. Based on the value of the scalar curvature  in a special gauge and the zero-flux equation the value of the ``cyclotron" spin as a non-universal quantity  in this 3-band case is inferred.
Section~\ref{Conclusions} is devoted to conclusions.

 \section{Berry curvature and gauge transformations}
\label{Berry curvature and gauge transformations}

In this section the meaning and the form of the gauge transformations in the
context of the Bloch problem will be explained. We will closely follow the notation of Ref. \cite{pssH}.

Let the periodic crystal tight-binding Hamiltonian be
\begin{equation}
{\cal H} = \sum_{j,k} t_{jk} c_j^+ c_k,
\end{equation}
where $j$ or $k$ is a shorthand notations for a site in a crystal, $j \equiv \mathbf{R} + \mathbf{a}_j ($ or $ k \equiv \mathbf{R} + \mathbf{a}_k$).  Vectors, $\mathbf{R}+ \mathbf{a}_i$, describe the positions of the sites where $\mathbf{R}$ is the vector
of a particular unit cell, and $\mathbf{a}_i, i = 1,\ldots, n - 1$ is the relative position of an atom $i$ inside the unit cell with respect to the one at $\mathbf{R}$. Note that the vectors $\mathbf{a}_j$ are not unique, that is we can define different embeddings of atoms in the unit cell.

We diagonalize the Hamiltonian in the inverse space with Bloch eigenvectors,
\begin{equation}
|\Psi_n(k)\rangle = \sum_{\mathbf{R},j} u_{jn}(k) \exp\{\mathrm{i}(\mathbf{R}+\mathbf{a}_j)\mathbf{k}\} |\mathbf{R}, j\rangle,
\end{equation}
corresponding to eigenvalues $\epsilon_n(k)$,
\begin{equation}
{\cal H} |\Psi_n(k)\rangle = \epsilon_n(k) |\Psi_n(k)\rangle.
\end{equation}
In order to define Berry curvature we define the embedding operator,
\begin{equation}
U^{\text{em}}(k) = \sum_{\mathbf{R},j} \exp\{\mathrm{i}\mathbf{k}(\mathbf{R}+\mathbf{a}_j)\}|\mathbf{R}, j\rangle\langle\mathbf{R}, j|,
\end{equation}
and its action on the Bloch vector as
\begin{equation}
|\Phi_n(k)\rangle = U^{\text{em}}(-k) |\Psi_n(k)\rangle.
\end{equation}
The Berry curvature is
\begin{align}
&B_n(k) = \nonumber \\
&- \mathrm{i} (\langle\partial_{k_x}\Phi_n(k)|\partial_{k_y}\Phi_n(k)\rangle -
\langle\partial_{k_y}\Phi_n(k)|\partial_{k_x}\Phi_n(k)\rangle ).
\end{align}
Because there is always more than one way to embed the atoms, i.e. fix coordinates of the atoms in the unit cell, the $u_{in}$'s are defined up to the shifts by the vectors of the unit cell, i.e.
\begin{equation}
u_{jn}(k) \rightarrow u_{jn}(k) \exp\{\mathrm{i}\mathbf{k}(\mathbf{a}_j - \mathbf{a}_j^{'} )\}, \label{shifts}
\end{equation}
i.e. by the vectors $ \mathbf{a}_j - \mathbf{a}_j^{'} = \mathbf{b}_j , j = 1,\ldots, n - 1$ - the vectors of a unit cell.

Another way to write the Bloch vector is
\begin{equation}
|\Psi_n(k)\rangle = \sum_j u_{nj}(k) c^+_{kj} |0\rangle
\end{equation}
where we defined a state,
\begin{equation}
 c^+_{kj} |0\rangle = \sum_{\mathbf{R}} \exp\{\mathrm{i}\mathbf{k}(\mathbf{R}+\mathbf{a}_j)\}|\mathbf{R}, j\rangle.
\end{equation}
As a quantum mechanical state it is defined up to a phase,
\begin{equation}
 c^+_{kj} \rightarrow \exp\{\mathrm{i} \alpha_j(k)\} c^+_{kj},
\end{equation}
and therefore we have a freedom in choosing $u_{nj}(k)$:
\begin{equation}
u_{nj}(k) \rightarrow \exp\{ - \mathrm{i} \alpha_j(k) + \mathrm{i} \alpha(k) \} u_{nj}(k).
\end{equation}
Here $\alpha(k)$ is the usual phase of the $U(1)$ transformation. We consider the phases,
$\alpha_j(k)$'s, to be analytic periodic functions in inverse space,
\begin{equation}
\alpha_j(k) = \alpha_j(k + K) + \mathrm{mod}\; 2 \pi,
\label{periodic}
\end{equation}
where $\mathbf{K}$ is any vector of the inverse space, so that the matrix element,
\begin{eqnarray}
&&\langle\Psi(k_1)|\Psi(k_2)\rangle = \nonumber \\
&&\sum_{j,K} \exp\{-\mathrm{i} \mathbf{K} \mathbf{a}_j\} u_j(k_1)^{*} u_j(k_1 + K) \delta_{k_2,k_1 + K},
\end{eqnarray}
remains unchanged under gauge transformations.
In general, as we will show in examples,
the Berry curvature is not invariant under these transformations, which include
those described in the expression (\ref{shifts}). Nevertheless one may expect an existence of a ``physical gauge" which will respect the symmetries of the crystal \cite{preprint,rmp}, but we will not go into that
question further. Our program will be to find gauge invariant quantities and give their physical interpretation.

\begin{figure}[h]
\centering
\includegraphics[width=0.9\linewidth]{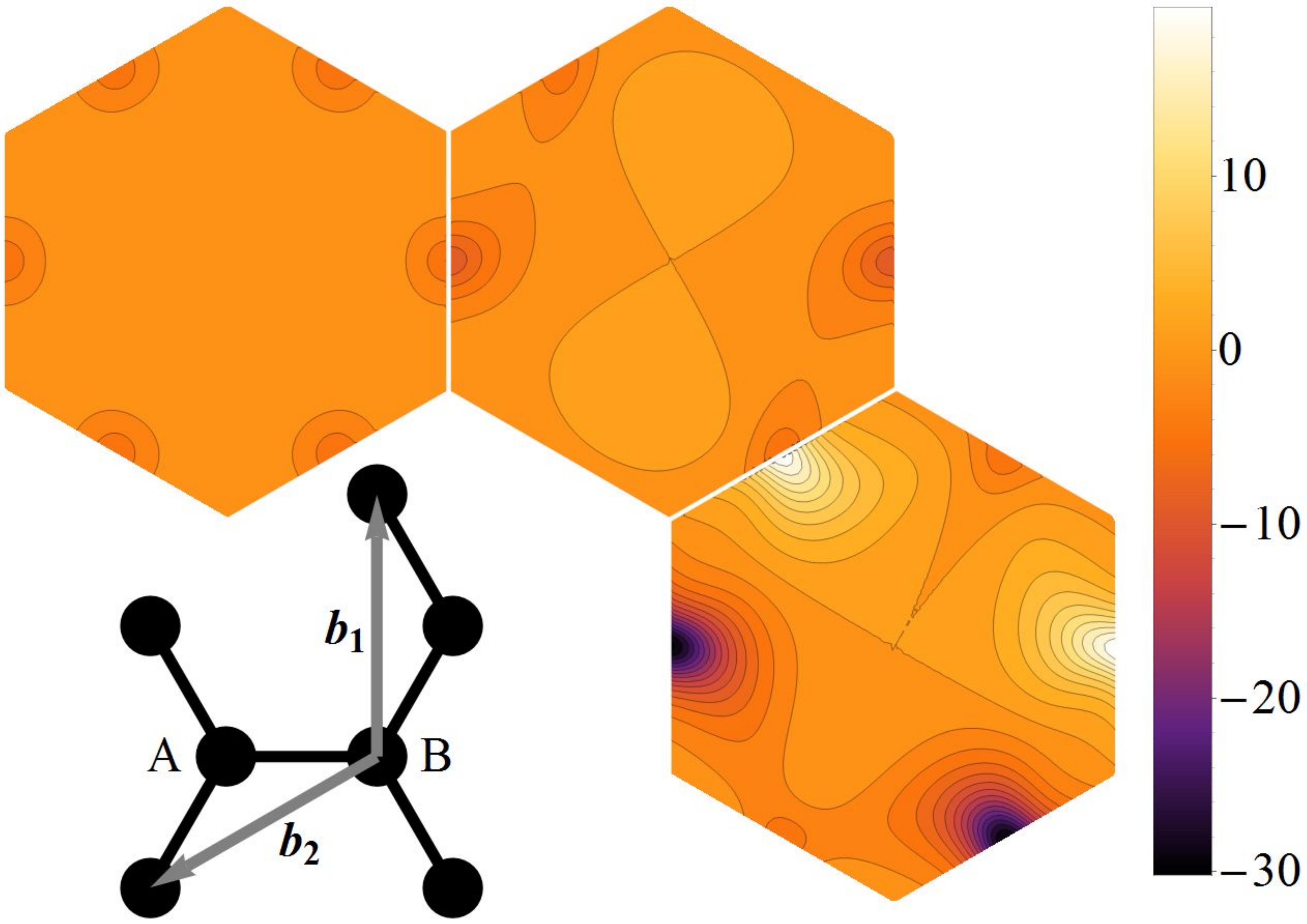}
\caption{\label{BerryC} The Berry curvature of the Haldane model in the Brillouin zone on the hexagons:
(up-left) in the case with no gauge transformation ($g(k)=I$),
(up-right) $g(k)=g_1(k)$, and
(down-right) $g(k)=g_2(k)$. Gauge transformations $g_1(k)$ and $g_2(k)$ are defined in the text.
Down-left is given part of graphene sheet with direct lattice vectors $\mathbf{b}_1$ and $\mathbf{b}_2$, and sublattices A and B.}
\end{figure}

In Fig. \ref{BerryC} the Berry curvature of the Haldane model \cite{Haldane_parity} for the value, $\varphi = 0.125\pi$, of the phase of the complex hopping between second neighbors,  is shown for three different gauges: in the case with the Hamiltonian of Ref. \cite{Haldane_parity} and eigenstate $u_k$ non-periodic in the inverse space, and two other states of the form
$ g(k)\; u_k $ , where $g(k)$ is a non-trivial gauge transformation. We define $g(k)$ as diagonal gauge transformations with the following form:
\begin{equation}
g(k) = \begin{bmatrix}
\exp\{\mathrm{i} \alpha_{1}(\mathbf{k})\} & 0 \\
0 & \exp\{ \mathrm{i} \alpha_{2}(\mathbf{k})\}
\end{bmatrix},
\end{equation}
where $\alpha_{1}(\mathbf{k})$ and $\alpha_{2}(\mathbf{k})$ are analytic, periodic in the inverse space (Eq.(\ref{periodic})), and everywhere well defined functions of $\mathbf{k}$.
In the second case (unit cell reparametrization)
$g(k)$ is
 \begin{equation}
g_1(k)  = \left[\begin{array}{cc}
1 & 0 \\ 0 & \exp\{\mathrm{i} \mathbf{k}\mathbf{b}_2\}
\end{array} \right],  \label{vectorH}
\end{equation}
while in the third case,
\begin{equation}
g_2(k)  = \left[\begin{array}{cc}
1 & 0 \\ 0 & \exp\{\mathrm{i} \left(\mathbf{k}(\mathbf{b}_1+\mathbf{b}_2)\right)^2\}
\end{array} \right].  \label{vectorH}
\end{equation}
Here vectors, $\mathbf{b}_j; j = 1,2$, of the direct lattice are shown and defined in the same figure. In all cases the Chern number, $C = -1$, is the same. But what is also remarkable is that the value of the Berry curvature at the high-symmetry point $K$ stays the same. We will discuss this more in the following.

\section{Dirac based 2-band CIs and introduction to the geometric description of the single particle problem}
\label{Dirac based 2 band CIs and the geometric description of the single particle problem}

In the usual gauges~\cite{Haldane_parity,Neupert}
the Hamiltonian of the Haldane model around $K$ point (the graphene expansion point) is
\begin{equation}
{\cal H}_{k} =  \begin{bmatrix}
- p  & v (k_x - \mathrm{i} k_y) \\
v (k_x + \mathrm{i} k_y) & p
\end{bmatrix}, \label{H}
\end{equation}
where $k = k_x + \mathrm{i} k_y$ is the complex Bloch momentum, $v = (\sqrt{3}/2) t_1 a$, $ p = 3 \sqrt{3} \sin(\varphi) t_2$,
with the first neighbor hopping parameter $t_1$ and the second neighbor hopping parameter $t_2$, the distance between second neighbors on the honeycomb lattice $a$, and the phase of the complex hopping between second neighbors $\varphi$. To the second order in $k$ we  find that the normalized Bloch state of the lower band effectively, i.e.
in the long distance is
\begin{equation}
u_k = \left[\begin{array}{c}
1 - l_D^2 \frac{|k|^2}{2} \\ - l_D k
\end{array} \right],  \label{vectorH}
\end{equation}
where $l_D = v/2p$. The Berry curvature at point $K$ is
\begin{equation}
B(K) = - \mathrm{i} (\partial_{k_x} u^+_k \partial_{k_y} u_k - \partial_{k_y} u^+_k \partial_{k_x} u_k ) = 2 l_D^2. \label{bc}
\end{equation}
The Berry field characterizes the change of phase of the Bloch state inside the Brillouin zone.
Here we notice two characteristic lengths, $l_D$ and $l_B = \sqrt{B(K)}$.  We notice that the substitution  $l_D^2 = l_B^2/2$  in~(\ref{vectorH}) will lead to the Gaussians of the lowest Landau level (LLL), $ (1 - l_D^2 |k|^2/2 \approx \exp\{- l_B^2 |k|^2/4\})$ in inverse space.

One can check that the value of Berry curvature at $K$ point is invariant under the gauge transformations, $u_k \rightarrow g(k) u_k$, where
\begin{equation}
g(k) = \begin{bmatrix}
\exp\{\mathrm{i} \alpha_{1}(\mathbf{k})\} & 0 \\
0 & \exp\{ \mathrm{i} \alpha_{2}(\mathbf{k})\}
\end{bmatrix}, \label{gauge_transformation}
\end{equation}
and $\alpha_{1}(\mathbf{k})$ and $\alpha_{2}(\mathbf{k})$ are analytic, everywhere well defined functions of $\mathbf{k}$.

The Berry curvature at the other, time-reversed, $K$ point is infinite, and this point cannot be an expansion point
for a geometrical description of an interacting problem. This point is isolated and does not contribute to the integral for the Chern number. With singular $U(1)$ gauge transformations like $k^*/|k|$ we can reverse the values and behavior of $K$ points.

The corresponding quantity that characterizes the change of amplitude of the Bloch state is Fubini-Study metric:
\begin{align}
& g_{ij}^{FS}(k) = \frac{1}{2}  [ \partial_i u_{\alpha_1,k} \partial_j u_{\alpha_1,k}^{*} + \partial_j u_{\alpha_1,k} \partial_i u_{\alpha_1,k}^{*} -\nonumber \\
& \partial_i u_{\alpha_1,k} u_{\alpha_1,k}^{*} u_{\alpha_2,k}\partial_j u_{\alpha_2,k}^{*} - \partial_j u_{\alpha_1,k} u_{\alpha_1,k}^{*} u_{\alpha_2,k}\partial_i u_{\alpha_2,k}^{*}]. \label{fs}
\end{align}
Indexes $\alpha_1$ and $\alpha_2$ label the Bloch state
components, and we assumed the summation over the repeated indexes.

The space of Bloch vectors is $ CP^1$, i.e. a space of vectors which are normalized and have $2$ complex components. In such a space, with metric $g_{ij}^{FS}(k)\equiv g_{ij}$, the affine connection is given by
\begin{equation}
\Gamma^k_{ij} = \frac{1}{2} g^{km} (g_{mi,j} + g_{mj,i} - g_{ij,m}).
\end{equation}
The connection defines the curvature tensor $R^i_{jkl}$ and the scalar curvature $R = g^{jl} R^i_{jil}$. Inserting~(\ref{fs}) for the $g_{ij}$, we find that in the case of Haldane model $ R = 8$ across the Brillouin zone~\cite{f1}.
This is in agreement with the expectation of the behavior of the scalar curvature in $CP^n$, $n = 1$.

Let us introduce
\begin{equation}
w_i^{FS} = \frac{1}{\sqrt{2}} \left[\partial_i u - (u^+ \partial_i u) u\right] \equiv \frac{1}{\sqrt{2}} D_i u. \label{Bloch_zweibein}
\end{equation}
The $u$ denotes a Bloch two-component vector, and we suppressed the orbital index $\alpha$, $\alpha = 1,2$ in $ w_{i,\alpha}^{FS}$.
In terms of $w_i^{FS}$ the Berry curvature is
\begin{equation}
B = 2 (-\mathrm{i})\left(w_i^{FS+} w_j^{FS} - w_j^{FS+} w_i^{FS}\right),
\end{equation}
and the Fubini-Study metric is
\begin{equation}
g_{ij}^{FS} = w_i^{FS+} w_j^{FS} + w_j^{FS+} w_i^{FS}.
\end{equation}
Therefore the $ w_{i}^{FS}$'s play the role of complex vielbeins that is  zweibeins in 2 dimensions.
Due to the relationship between Berry curvature and (the square root of) the determinant of the FS metric, $ \det(g_{ij}^{FS}) \equiv g^{FS}$,
\begin{equation}
\frac{B}{2} = \sqrt{g^{FS}},
\end{equation}
valid in the whole BZ in the case of 2-band models, we can express the  local spin connection, $\Omega_i^{FS}$, and curvature $ R^{FS} = \epsilon_{ij} \partial_i \Omega_j^{FS}$, in terms of the complex zweibeins of the FS metric, Eq.~(\ref{Bloch_zweibein}).
The derivation and expressions can be found in Appendix. The result for the spin connection is
\begin{equation}
\Omega_i = - \frac{1}{\sqrt{g}} \epsilon^{kl} (\partial_i w_k^+) w_l +
\frac{1}{2 \sqrt{g}} \epsilon^{kl} \partial_k g_{il} + \frac{\mathrm{i}}{4} \partial_i (\ln g), \label{sc}
\end{equation}
with  $w_i = w_{i,\alpha}^{FS}$ and $g_{ij} = g_{ij}^{FS}$ this becomes $\Omega_{i}^{FS}$.
The expression~(\ref{sc}) is also the spin connection for a general non-uni-modular metric for which
\begin{equation}
\frac{\tilde{B}}{2} = \sqrt{g},
\end{equation}
and
\begin{equation}
g_{ij} = w_i^+ w_j + w_j^+ w_i,
\end{equation}
and
\begin{equation}
\tilde{B} = 2(-\mathrm{i})(w_i^+ w_j - w_j^+ w_i),
\end{equation}
holds.
The expression for the temporal part of the spin connection is
\begin{equation}
\Omega_0  =  - \frac{1}{\sqrt{g}} \epsilon^{kl} (\partial_0 w_k^+) w_l + \frac{\mathrm{i}}{2} \partial_0 \ln g,
\end{equation}
where we assumed $ g_{i0} = 0, i = 1,2$, and time dependent zweibeins.

\section{The geometric description of (F)QHE - overview}
\label{The geometric description of (F)QHE - overview}
In the case of (F)QHE the kinetic part of the Lagrangian with geometric degrees of freedom is described
in~\cite{Haldane_talk,Maci,YeJe}. In Ref.~\cite{Maci} the nematic degrees of freedom act as geometric in the description of the FQHE.  The part of the Lagrangian that is linear in time derivative is
\begin{equation}
{\cal L}^{n} = s \rho \Omega^n_0 + \cdots, \label{nematic_l}
\end{equation}
where $s$ is the total spin of the topological state with both cyclotron and guiding center contribution. The density $\rho$ is given by  $\rho = \nu/2 \pi l_B^2$  where $\nu$ is the filling factor and $l_B$ is the magnetic length. The time component of the spin connection, $ \Omega_0 = \epsilon^{ij} z^*_i \partial_0 z_j$, comes from the geometric, i.e. nematic degree of freedom described by the uni-modular metric
$g_{ij}^n = z_a^* z_b + z_b^* z_a$. This uni-modular metric is a matrix exponential, $\hat{g}^n = \exp \hat{Q}$, where $\hat{Q}$ is the traceless, geometric nematic matrix order parameter.  In the same reference the well-known expression for the Hall viscosity in (F)QHE~\cite{Avron,rr},
$ \eta_H^{FQHE} = s \rho/2$, was derived on the basis of the nematic description.

On the other hand, in the geometric description of FQHE introduced in Ref.~\cite{Haldane_geo_prl,Haldane_talk} only guiding center metric is the dynamical degree of freedom. The resulting term in the Lagrangian is
\begin{equation}
{\cal L}^{gc} = \bar{s} \rho \Omega_0 + \cdots,\label{geo_l}
\end{equation}
where $\bar{s}$ is the guiding center spin. For the ideal Laughlin state $\bar{s}$ is equal to $\bar{s}= (m - 1)/2$, while in Eq.~(\ref{nematic_l}) $s$ is equal to $s = m/2$. The expression~(\ref{geo_l}) can be derived considering the Wen-Zee form~\cite{wz} of the Chern-Simons (CS) action on curved spaces with the introduction of the dynamical spin connection which couples to the guiding center (particle) current~\cite{YeJe}.

\section{The geometric description of the ground state of non-interacting and interacting Dirac based 2-band CIs}
\label{The geometric description of the ground state of interacting Dirac based CIs}
Due to the presence of the (constant) curvature in the space of Bloch vectors for 2 -band CIs we may expect an additional flux just as in the continuum case. The equation of state of the single particle problem (non-interacting Chern insulator) (expressed in the form of quadratic Hamiltonians) is
\begin{equation}
B(k) - s \sqrt{g^{FS}} \frac{R^{FS}}{2} = 0, \label{nonint_eq}
\end{equation}
because $R^{FS} = 8$, and we take  $ s = 1/2$. Using the spin connection~(\ref{sc}) we can also write~(\ref{nonint_eq}) as
\begin{equation}
B(k) - s \epsilon^{ij} \partial_i \Omega_j^{FS}(k) = 0. \label{nonint_eq_sc}
\end{equation}
This equation tells us that the total local flux in BZ experienced by  particles  is zero.

In the quantum mechanical description of the non-interacting particle in 2D in the presence of perpendicular to the plane magnetic field, there is no difference between coordinate and momentum representation in the rotational symmetric gauge. The only exception is  the place of the dimensional factor, $(\mathrm{ magnetic \; length})^2$, in the Gaussians of the lowest Landau level. Therefore, at least in  the case with the rotational symmetry around a high-symmetry point, we expect that the form of the appropriate effective Wen-Zee action in the (fractional)CI case is the same as in the FQHE case with the indexes referring to $(\mathbf{k}, t)$, momentum - time instead of $(\mathbf{r}, t)$, space - time.
Therefore we will assume in the following that the effective description of fractional CIs
can be given by the same form of the Wen-Zee Lagrangian in the inverse space of Bloch vectors. The point of the effective description should coincide with the point of the smallest gap, i.e. the low-energy description. In this way we have chosen the expansion point in the usual way for a condensed matter system as the point of low-energy description. But potentially (as we will see in the two examples of 2 and 3 band CIs) in the case of the projection to flat bands, the expansion points may be defined as points of gauge-invariant Berry curvature solely.
In a basic description of the Lagrangian we omit terms that specify energetics. These terms
will differentiate between these two cases: with and without (flat)
energy dispersion, but the basic response and form of the Lagrangian will not differ.

Therefore, we take~(\ref{nonint_eq_sc}) as the equation of the single particle problem and introduce the dynamical spin connection,
\begin{equation}
\Omega_i(k,t) = \Omega_i^{FS}(k) + \delta \Omega_i (k,t),
\end{equation}
with $ \delta \Omega (k)$ due to interactions.
In terms of zweibeins we have
\begin{equation}
\Omega(k,t) = \tilde{\Omega}(w^{FS} + \delta w),\; \text{with } \tilde{\Omega} (w^{FS}) = \Omega^{FS}.
\end{equation}
The Wen-Zee action for CIs is given  by
\begin{equation}
{\cal L} = \hbar \left( - \frac{\alpha \partial \alpha}{4 \pi} - \frac{[\partial A - s_r \partial \Omega ]\alpha}{2 \pi}\right), \label{lagrangian_ci}
\end{equation}
with a shorthand notation for a field $\beta_\mu$: $
\epsilon_{\lambda \mu \nu} \partial_\mu \beta_\nu \equiv \partial \beta. $
In the Lagrangian $s_r = s + \delta s = 1/2 + \delta s$, with $\delta s$ due to interactions, and $A_\mu$ represents the static Berry connection  with $A_0 = 0$.
The particle density-current is given by
\begin{equation}
j_\mu =\frac{1}{2 \pi} \partial \alpha,
\end{equation}
and the classical equation of motion for field $\alpha$ is
\begin{equation}
j_\mu =\frac{1}{2 \pi} \left[\partial A - s_r \partial \Omega \right] = - \delta s \partial\Omega^{FS} - s_r \partial\delta\Omega.
\end{equation}
Here $j_0$ represents departure from the uniform density in the ground state. Therefore the change in the particle density comes from both, cyclotron (single particle background) and guiding center (interacting) degrees of freedom through their interference.
After integrating out $\alpha$ (particle degree of freedom) we obtain
\begin{equation}
{\cal L} = \frac{\hbar}{4 \pi}  \left[ A - s_r  \Omega \right]\left[\partial A - s_r \partial \Omega \right].
\end{equation}
The part of the Lagrangian linear in time derivative is
\begin{align}
{\cal L} &= \frac{\hbar}{4 \pi}  \left( - 2 s_r B \Omega_0 + 2 s_r^2 \epsilon_{ij} \partial_i \Omega_j \Omega_0 + s_r^2 \epsilon_{ij} \Omega_i \partial_0 \Omega_j\right) \nonumber \\
&= \frac{\hbar}{2 \pi} \delta s B \Omega_0 + \cdots \label{top_part}
\end{align}
In the last line we approximated $(2 s_r) \delta s \approx \delta s$ and $ s \; \epsilon_{ij} \partial_i \Omega_j \approx B$.

Before discussing Eq.~(\ref{top_part}) in detail let us make a few comments.
First we would like to point out that we are describing 2 band problem. The one band (i.e. interacting and filled band) problem is trivial as the density, i.e. occupation number, is constant. In the 2 band problem the mixing of ``cyclotron" and guiding center degrees of freedom is expected and this motivates the coupling we introduced in Eq.~(\ref{lagrangian_ci}). Also we are assuming that the system preserves, despite band mixing, the Hall conductance quantization. We justify this by an assumption that interactions act as small perturbation which do not change the Chern number quantization.

We should also comment that we used the same form of the spin connection, Eq.~(\ref{sc}), for the dynamical zweibein $w = w^{FS} + \delta w$ as for the Bloch vector based zweibein, Eq.~(\ref{Bloch_zweibein}).
Thus we assumed that at any momentum-time point $\tilde{B}/2 = \sqrt{g}$ holds (see Section~\ref{Dirac based 2 band CIs and the geometric description of the single particle problem}), where $\tilde{B}$ is the dynamical Berry field, based on the dynamical $w$. This assumption can be stated differently as a demand that for the dynamical Berry curvature $\tilde{B}$ and connection $\tilde{A}$ the following requirement holds,
\begin{equation}
\int_{BZ} \tilde{B} \mathrm{d}^{2}k =\int_{BZ} \epsilon_{ij} \partial_i \times \tilde{A}_j \mathrm{d}^{2}k = 2 \pi C.
\end{equation}
Here $C$ is the Chern number equal to $C=1$ as in the case without the interactions. Then this can be solved (reduced over the $S^2$, i.e. two-sphere angle integration) by taking
$ \tilde{A}_j = (-\mathrm{i}) \tilde{u}^{*} \partial_j \tilde{u} $ where $\tilde{u}$ is the two-component complex normalized vector field, i.e. we can have
$w_{i,\alpha} = D_i \tilde{u}_{\alpha}/\sqrt{2}$.
Thus at any point $\tilde{B}/2 = \sqrt{g}$ holds and the expression for the spin connection, Eq.~(\ref{sc}), follows. Also, in this way, it follows that we consider the space of (dynamical) zweibeins which are smoothly connected to those based on Bloch vectors in the non-interacting problem.
\section{Discussion}
\label{Discussion}
In Eq.~(\ref{top_part}) the quantity $\delta s$  plays the role of the guiding center spin. If the same formalism is applied to the FQHE in the case of ideal Laughlin case we would have $ \delta s/2 \pi \rightarrow \delta s/2 \pi m =
(m - 1)/4 \pi m$, compare with Eq.~(\ref{geo_l}). In that case $A$ is the external vector potential of uniform magnetic field and $R^{FS}$ is zero.
If the rotational symmetry is assumed, the coefficient in the Lagrangian
$(\hbar/2 \pi m) \delta s B = (\hbar/2 \pi m) \delta s l_B^2$ in the inverse space becomes $(\hbar/2 \pi m) \delta s (1/l_B^2)$ in the ordinary space. The coefficient in the ordinary space can be rewritten as
$ \hbar \delta s \rho$ and enters the expression for the Hall viscosity of guiding centers in FQHE:
\begin{equation}
\eta_H^{FQHE}  = \frac{\hbar \delta s \rho}{2} = \hbar \delta s (\frac{1}{4 \pi m l_B^2}).
\end{equation}
In this way we might expect   fixing $m = 1$ that the same formula in the case of CIs holds. But first we should carefully examine and compare expressions for the spin connection,
\begin{equation}
\Omega_0 = \epsilon^{ij} \frac{w_i^* \partial_0 w_j}{\sqrt{g}},
\end{equation}
in the Lagrangian in Eq.~(\ref{top_part}) in both cases, FQHE and CI. We will study small deformations, i.e. small fluctuations from a flat ground state configuration. This will just serve as a way to detect the role of the characteristic length in the contribution to the Hall viscosity from the internal (interacting) degrees of freedom.

Namely in the context of FQHE, $w_i$ and $g$ are dimensionless quantities and $\det{g} = 1$ (uni-modular metric requirement). To study the geometry deformations we rewrite $\Omega_0$ in terms of zweibeins $e^1_i$ and $e^2_i$, $i = 1,2$ defined as
\begin{equation}
e^1_i = \frac{1}{\sqrt{2}} (w_i + w_i^*) \;\; \textrm{and} \;\;  e^2_i = \frac{i}{\sqrt{2}} (w_i - w_i^*).
\end{equation}
The resulting expression is
\begin{equation}
\Omega_0 = e^1_1 \partial_0 e^1_2 - e^2_2 \partial_0 e^2_1.
\end{equation}
Then we  study deformations around a flat configuration, $ e^i_j = \delta^i_j$, encoded in the following
\begin{equation}
e^1_1 = 1 + e_1, e^2_2 = 1 - e_1, e^1_2 = e^2_1  =  e_2,
\end{equation}
with $e_1$ and $e_2$ small deformation parameters.
We find
\begin{equation}
\Omega_0 = 2 e_1 \partial_0 e_2.
\end{equation}

On the other hand, in the case of 2 band CIs, zweibeins have two  components denoted by $\alpha$ in $w_{i,\alpha}$. Thus  $ e^i_j \rightarrow e^i_{j \alpha}$ and can be easily found in the long-distance approximation in the Dirac case, Eqs.~(\ref{H}) and~(\ref{vectorH}):
\begin{equation}
e_1^1 = l_D  \begin{bmatrix}
1 \\ 0
\end{bmatrix},
e_2^2 = l_D \begin{bmatrix}
- 1 \\ 0
\end{bmatrix},
e_2^1 = e_1^2 = \begin{bmatrix}
0 \\ 0
\end{bmatrix}.
\end{equation}
We induce deformations as
\begin{equation}
e_1^1 =  l_D\begin{bmatrix}
e_1+ 1 \\ e_1
\end{bmatrix},
e_2^2 =   l_D\begin{bmatrix}
e_1 -1\\ e_1
\end{bmatrix},
e_2^1 = -  e_1^2 =  l_D e_2 \begin{bmatrix}
1 \\ 1
\end{bmatrix},
\end{equation}
i.e. we make deformations (study fluctuations) equal in both sublattices (otherwise we would make a spin torque transformation~\cite{You}). Here $e_1$ and $e_2$ are small deformation parameters.
Taking that $ \sqrt{g} \approx l_D^2$ in the ground state near expansion point we have  for the spin connection,
\begin{equation}
\Omega_0 \approx \epsilon^{ij} \frac{w_i^* \partial_0 w_j}{l_D^2} = 4 e_1 \partial_0 e_2.
\end{equation}
Therefore in performing the same deformations we find that there is an extra 2 in the final expression for the spin connection and eventually Hall viscosity in the case of CIs.
The Hall viscosity of the internal degrees of freedom in the inverse space is
\begin{equation}
\eta_H^{\text{inv}} = \hbar \delta s (\frac{l_B^2}{2 \pi}),
\end{equation}
leading to the Hall viscosity in the ordinary space
\begin{equation}
\eta_H = \hbar \delta s (\frac{1}{2 \pi l_B^2}) = \hbar \delta s (\frac{1}{4 \pi l_D^2}).
\end{equation}
The expression is in the complete agreement with the conclusion of Hughes et al.~\cite{Hughes2}. They found that in the case of Dirac based CIs the role of the magnetic length is taken by the $l_D$ length in the expression for the Hall viscosity.
Thus we find that a geometric description of the Dirac based Haldane model is possible. That is also true for any Dirac based CI, i.e. CI which low-energy description is given by Eqs.~(\ref{H}) and~(\ref{vectorH}).

\section{ 2-band models in general}
\label{Comparison with other 2 band models}
\begin{figure}[h]
\centering
\includegraphics[width=0.9\linewidth]{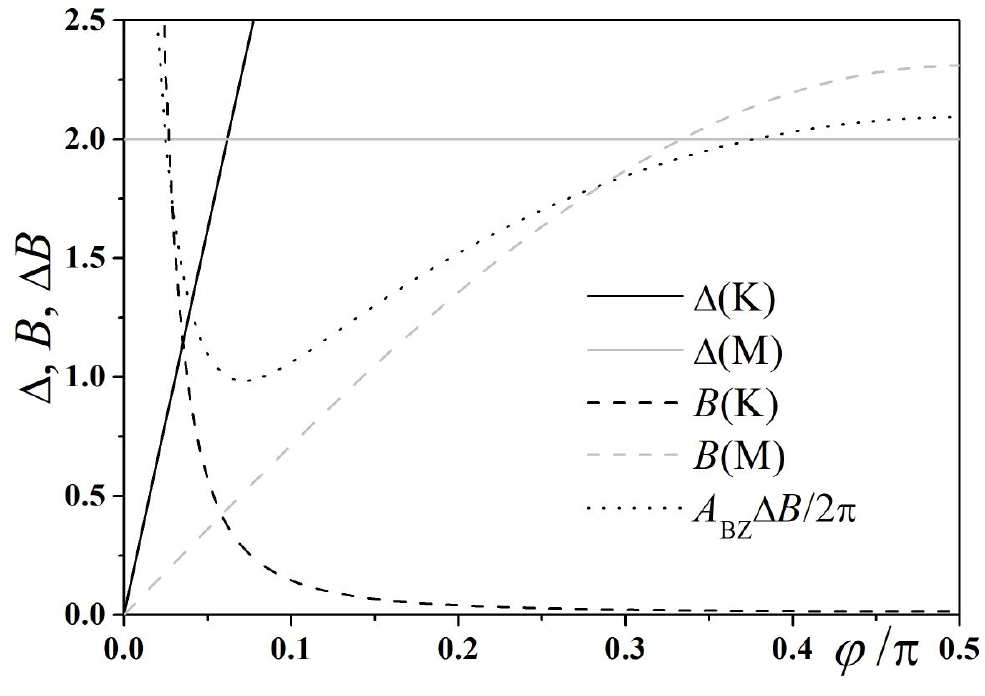}
\caption{\label{haldane} Berry curvature and direct gaps at points $K$ and $M$, and the dispersion of the Berry curvature as functions of $\varphi$. }
\end{figure}

The 2-band Haldane model is specific in having the property that we can recognize the effective cyclotron orbits in real space by looking at the Bloch vector in the expression~(\ref{vectorH}). Namely the weight on one of the sublattices is much larger (in the long distance limit) than on the other sublattice of the hexagonal lattice. Thus if we consider a superposition (wave packet) of Bloch vectors with Gaussians in $k$ space as cut-offs:
\begin{align}
u(z) &= \int \mathrm{d}\mathbf{k} \begin{bmatrix}
1 \\ - l_D k
\end{bmatrix} \exp\{- l_D^2 \frac{|\mathbf{k}|^2}{2}\} \exp\{\mathrm{i} (\mathbf{K}+ \mathbf{k}) \mathbf{r}\} = \nonumber \\
&= \exp\{\mathrm{i} \mathbf{K} \mathbf{r}\} \begin{bmatrix}
1 \\ - \frac{z}{l_D}
\end{bmatrix}\exp\{-  \frac{|z|^2}{2 l_D^2}\},
\end{align}
with coordinate $z = x + \mathrm{i} y$,
we recognize cyclotron orbit(s) with $l_B = l_D/\sqrt{2}$ as a characteristic length for the size of the orbit. If $l_B {\sim} a$, i.e. $l_D {\sim} a$ (lattice constant) we expect a QH like (LLL like) description of Haldane model based CI. Indeed small $\varphi$ (phase of complex hopping) for which  $l_B {\sim} a$ is the condition that characterizes a region of phase space of Haldane model (as shown in Ref.~\cite{DMR}) that is closest to the QH background and optimal for fractional CI physics. This region is closest to the effective (gapless) Dirac description and thus, it seems, recovering the Hall viscosity with $l_D$ playing the role of magnetic length is expected. Therefore the question is, when we lose the correspondence
$B(K){\sim}\bar{B}$ (average Berry curvature) or $l_D {\sim} a$, i.e. for $\sin(\varphi) {\sim} 1$,
whether the Dirac based description is still appropriate and geometric description is still possible.  Fig.~\ref{haldane} shows  Berry curvature and direct gaps at points $K$ and $M$, and the dispersion of the Berry curvature as functions of $\varphi$. We can see that with the increase of $\varphi$ the point $K$ becomes a point of larger
direct gap with respect to point $M$ though we are still in what we may call a QH region - the region of small Berry curvature dispersion. The point $M$ does not have  Berry curvature invariant under gauge transformations and thus it is not obvious candidate for the point of geometric description. Thus we may say that for larger
$\varphi$ the geometric description is not clearly defined.

On the other hand in the case of the 2-band based on quadratically dispersing Dirac model~\cite{Sun,You}
the Berry curvature at the expansion point $K$ is equal to zero. In this case, the geometric description of the ground state of CIs that we introduced in the Dirac based Haldane model is not possible. This makes the identification of the characteristic length difficult, especially in these models and bands with Chern number $|C| > 1$ that we will not consider further.
\section{3-band kagome model}
\label{3 band kagome model}
\begin{figure}[h]
\centering
\includegraphics[width=0.9\linewidth]{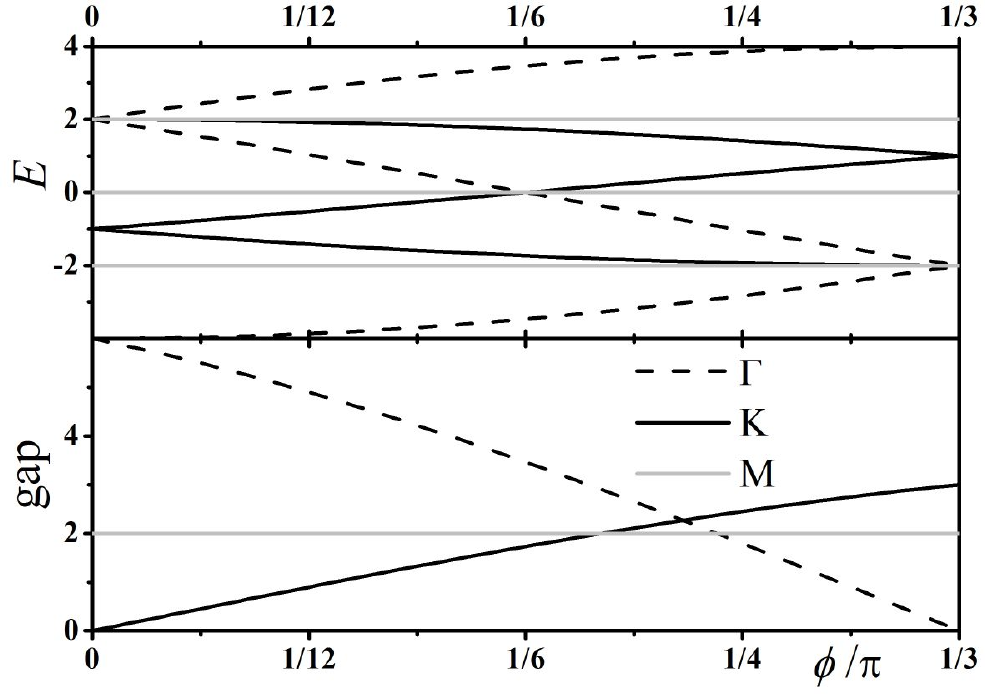}
\caption{\label{energies_kagome} Energies  and direct gaps between the lowest and middle band at points $K$, $M$, and $\Gamma$. }
\end{figure}

In this section we discuss whether the geometric description in the case of 3-band kagome model~\cite{TMW} is possible. The model is parametrized by the complex phase $\phi$ ($t_1+\mathrm{i}\lambda_1=t_1\exp\{\mathrm{i}\phi\}$ is a complex hopping parameter), and the system is gapless when $\phi = 0$ and $\phi = \pi/3$. The effective description at $\phi = 0$ is given around $K$ point in BZ (with linear Dirac-like dispersion) and at $\phi = \pi/3$ is given around  $\Gamma$ point. In Fig.~\ref{energies_kagome} we plotted the energies and direct gaps between the lowest and middle band at points $K$, $M$, and $\Gamma$.

\begin{figure}[h]
\centering
\includegraphics[width=0.9\linewidth]{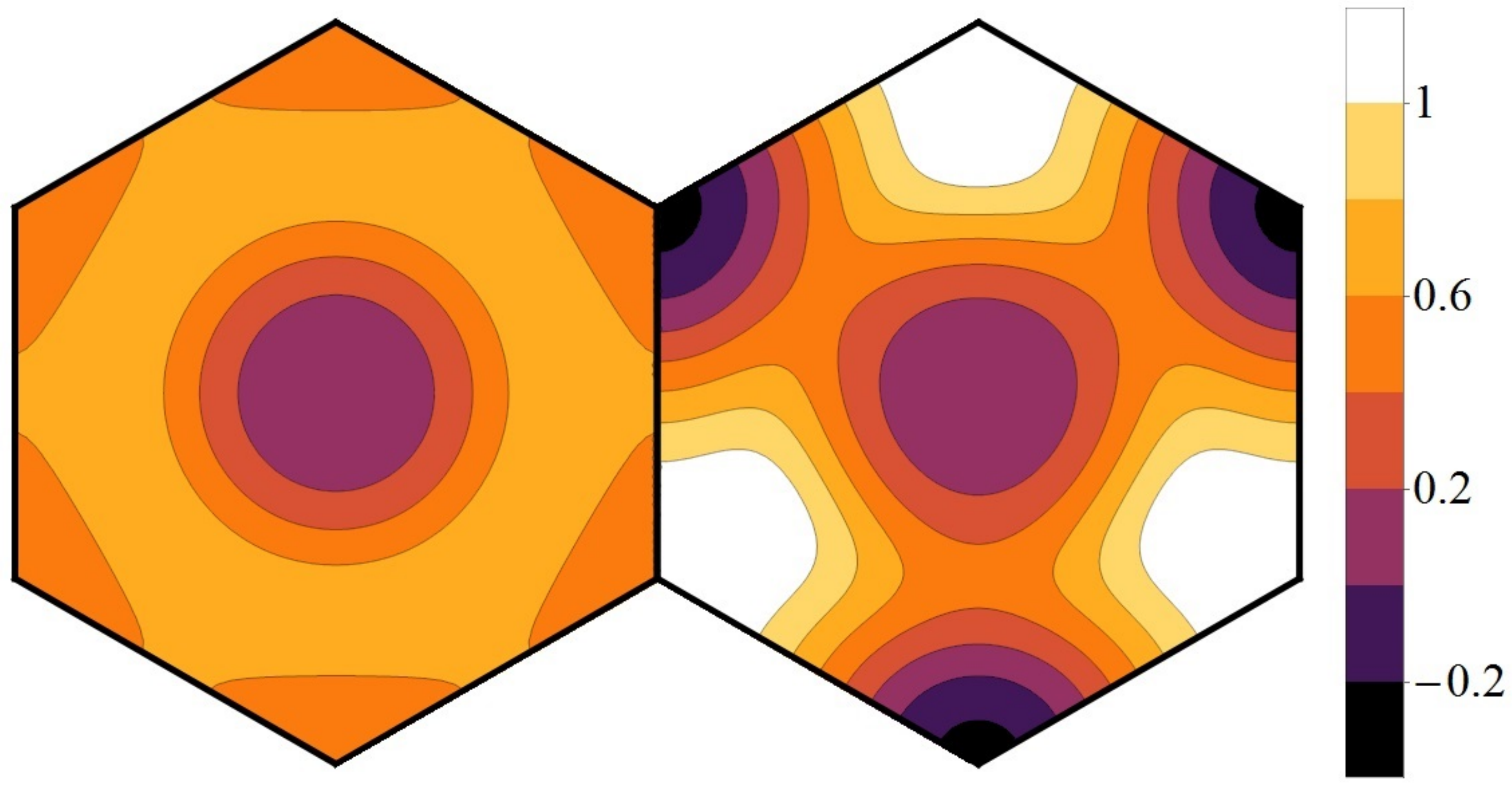}
\caption{\label{Fberry3band} Berry curvature in 3-band kagome model in TMW (left) and WBR (right) gauges.}
\end{figure}

\begin{figure}[h]
\centering
\includegraphics[width=0.45\linewidth]{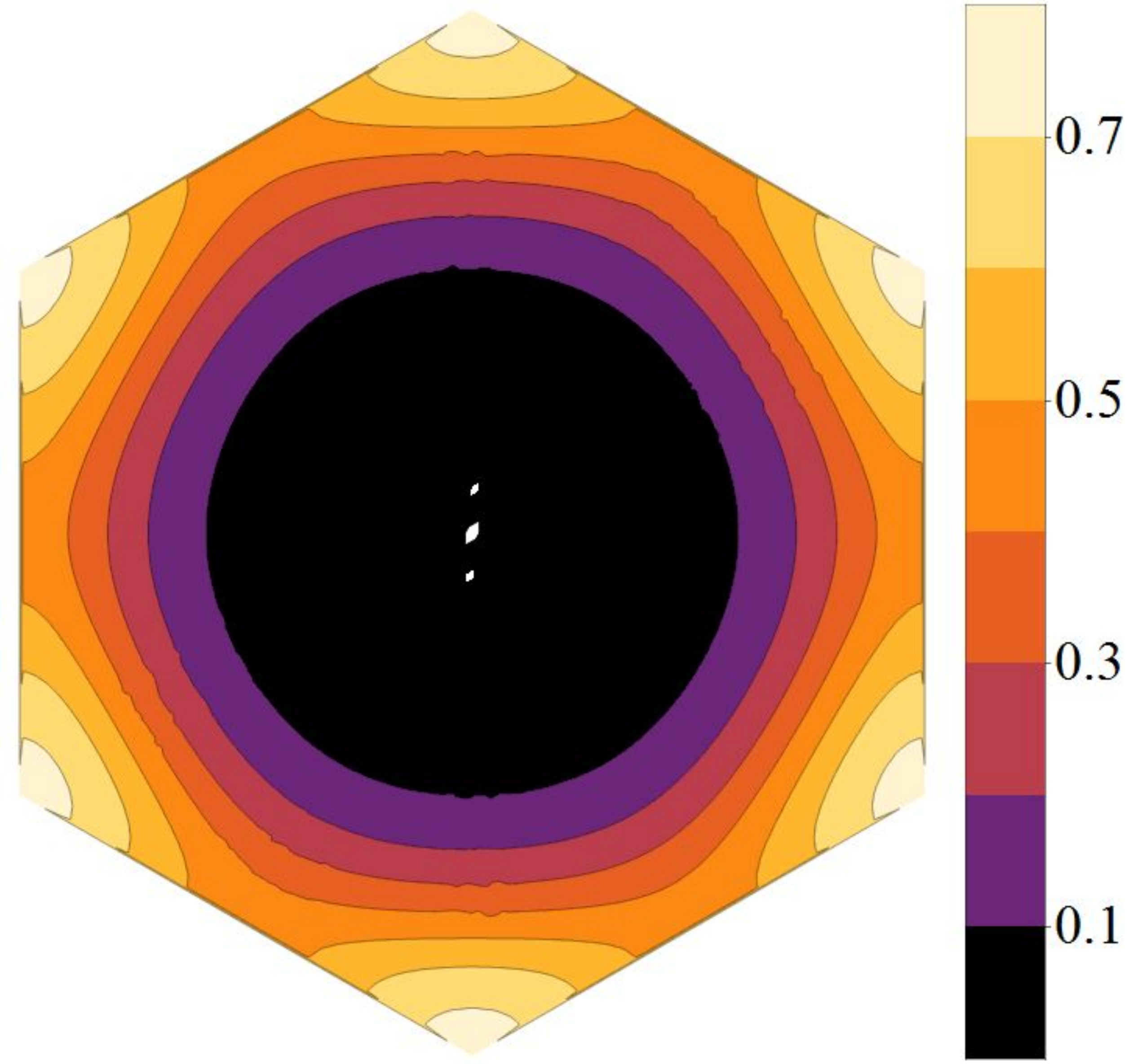}
\includegraphics[width=0.45\linewidth]{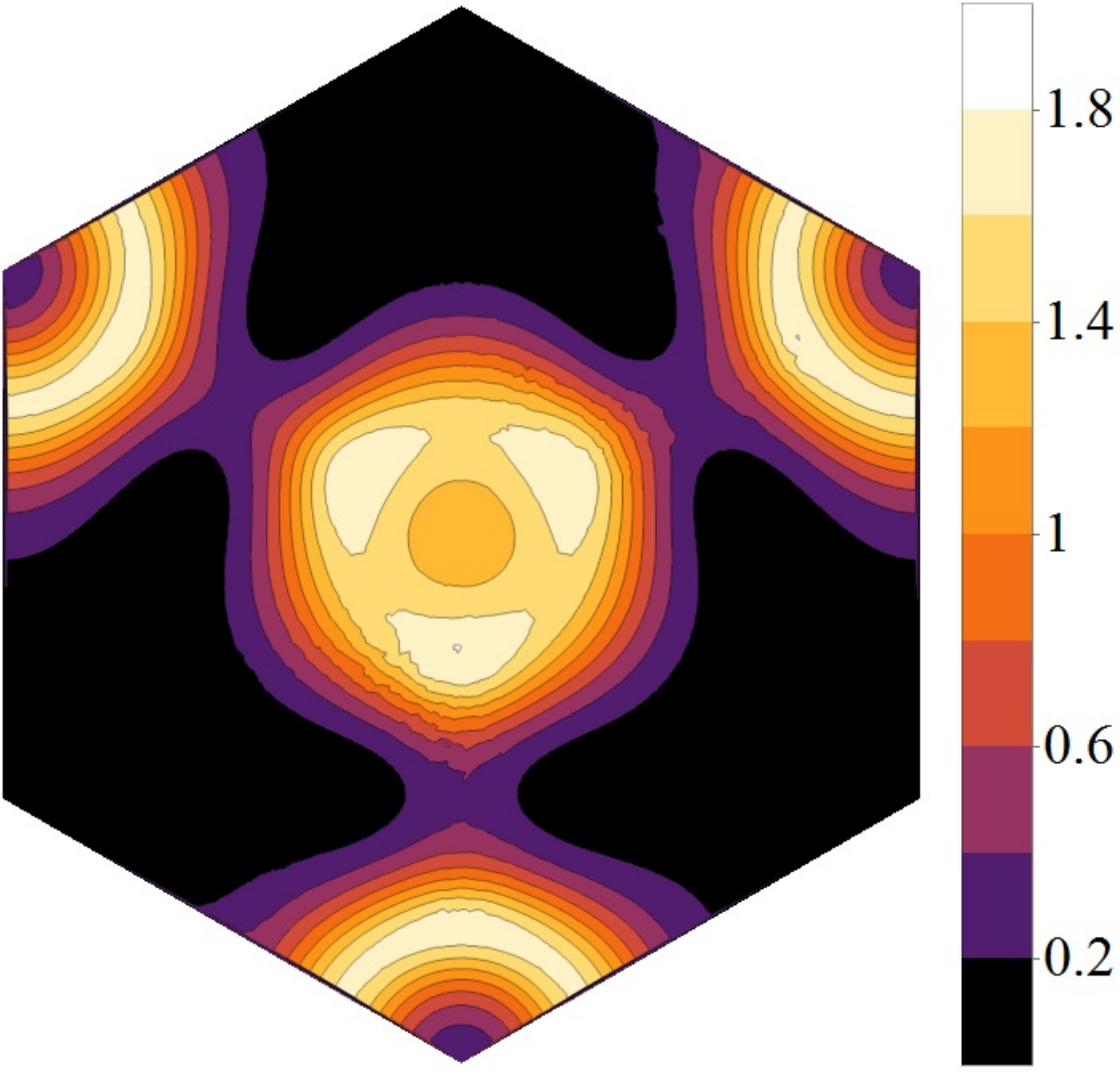}
\caption{\label{Fcurv3band} Left hand side of Eq.~(\ref{nonint_eq}) as function on Brillouin zone for 3-band kagome model in TMW (left) and WBR (right) gauges. In $K$ points in TMW gauge value is $\approx0.74$ while in WBR gauge values are different for $K$ ($\approx0.24$) and $K'$ ($\approx0.056$) points. Local spin $s(K)=2B/R\sqrt{g}$ that fulfills Eq.~(\ref{nonint_eq}) have values $\approx-0.85$ in TMW gauge, while in WBR gauge in $K$ points is $\approx0.26$ and have closest value to 1/2 in $K'$ points which is $\approx0.52$.}
\end{figure}
\begin{figure}[h]
\centering
\includegraphics[width=0.9\linewidth]{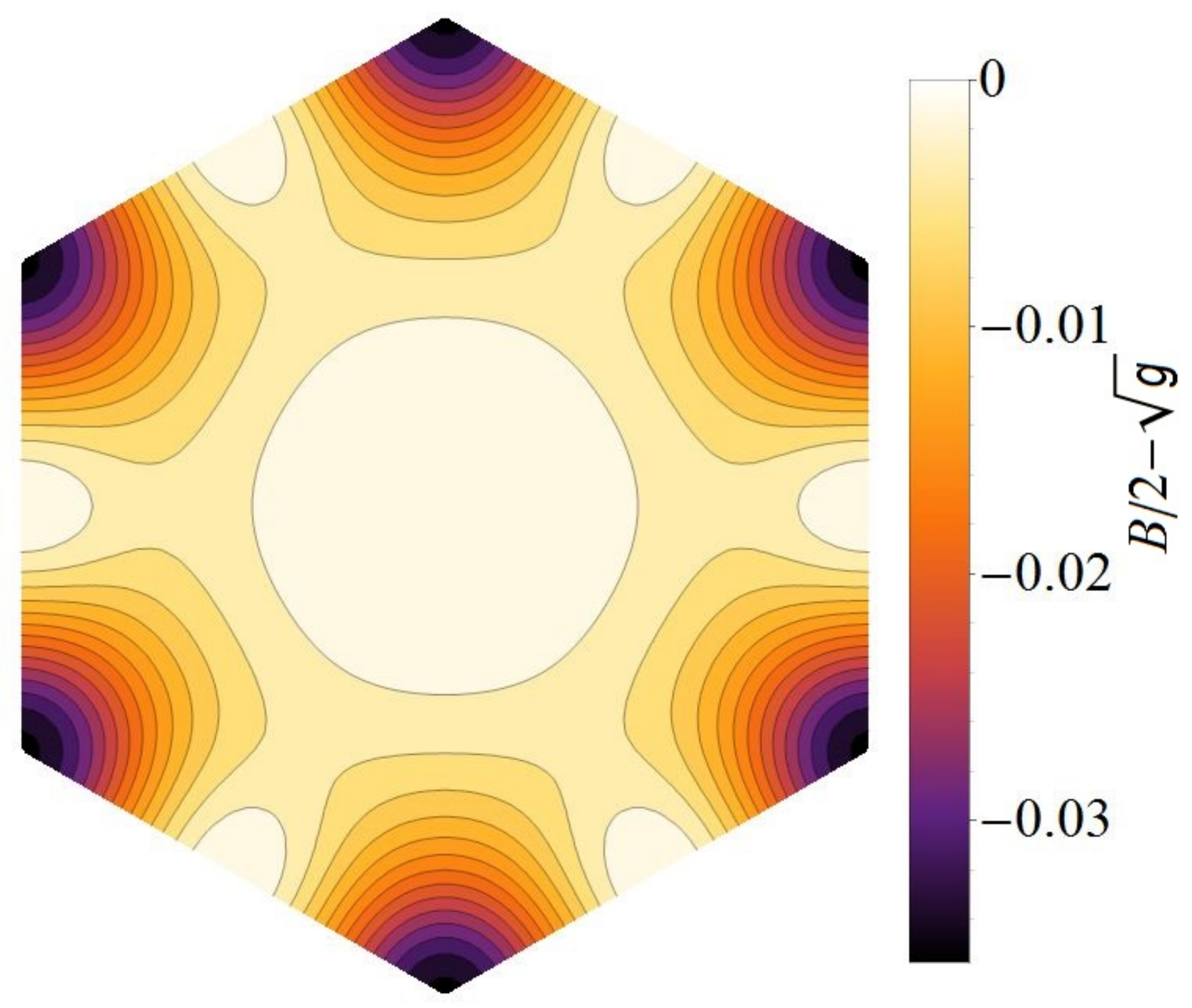}
\caption{\label{TMW_geo} The analysis whether in the gauge of Ref.~\cite{TMW} the equation $B/2 = \sqrt{g^{FS}}$ is fulfilled at $\phi = \pi/4$. Plotted is the difference $B/2 - \sqrt{g^{FS}}$.}
\end{figure}
\begin{figure}[h]
\centering
\includegraphics[width=0.9\linewidth]{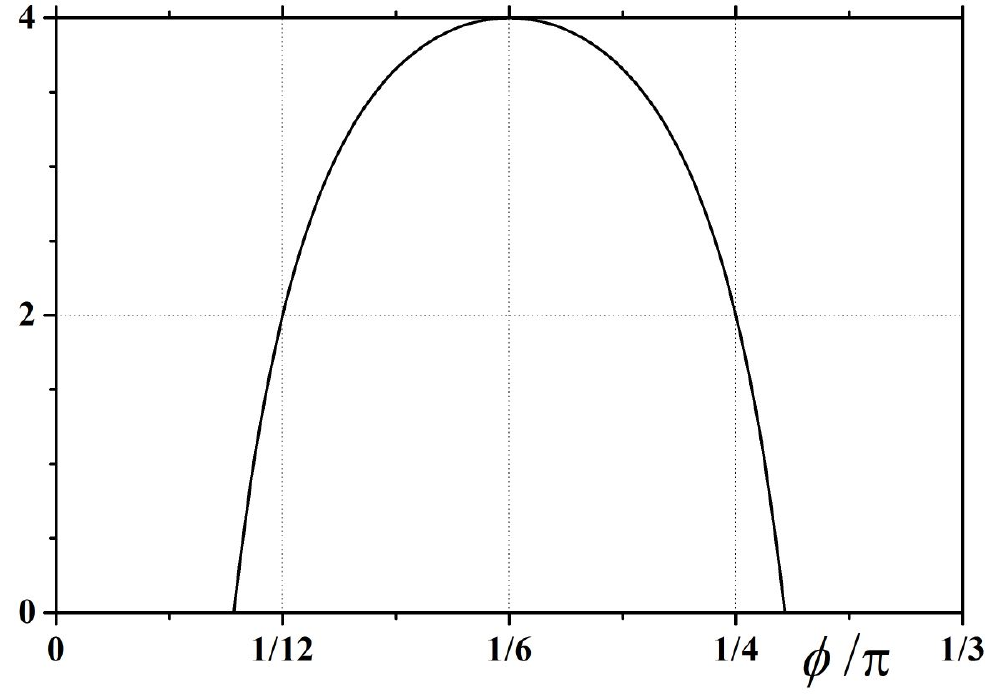}
\caption{\label{scalarRM}The scalar curvature, $R^{FS}$, in the special gauge of Ref.~\cite{TMW}, as a function of $\phi$, at $M$ points}
\end{figure}
We considered two gauges present in the literature, Ref.~\cite{TMW} and Ref.~\cite{zoo} with phase $\phi = \pi/4$, and examined the lowest lying band - a very good background for fractional CI states~\cite{zoo}. In Fig.~\ref{Fberry3band} the Berry curvature in BZ is plotted in these two gauges. We notice how very different graphs exemplify the fact that the Berry curvature is non-gauge invariant quantity. Nevertheless we used both gauges in the search for a universal expansion point as found in the case of the Haldane model. We looked whether the equation~(\ref{nonint_eq}) is fulfilled in BZ in both gauges and results are presented in Fig.~\ref{Fcurv3band}. Though point $K$ in the gauge of
Ref.~\cite{zoo} is very close to the fulfillment we could not find an appropriate expansion point for the geometric description.

Nevertheless, we notice in Fig.~\ref{Fberry3band} that the Berry curvature at point $M$
may be a gauge-invariant quantity. A close inspection shows that this is the case at  $M$ points for which the Berry curvature acquires the same, finite value ($ {\sim} \sin(\phi)$) that is invariant under diagonal gauge transformations with $g(\mathbf{k}) = \mathrm{diag}(\exp\{\mathrm{i} \varphi_{1}(\mathbf{k})\},\exp\{\mathrm{i} \varphi_{2}(\mathbf{k})\},\exp\{\mathrm{i} \varphi_{3}(\mathbf{k})\} )$ in the whole interval $\phi \in (0, \pi/3)$.  These $M$ points may serve as expansion points for the geometrical
description especially because at and in the neighborhood of these points the relationship $B/2 = \sqrt{g^{FS}}$ holds for any $\phi \in (0, \pi/3)$ in the gauge of Ref.~\cite{TMW}. In Fig.~\ref{TMW_geo} one can see an illustration of this in the case $\phi = \pi/4$. Thus, at these $M$ points two main assumptions of the geometric program (1) gauge invariant Berry curvature and (2) the equation $B/2 = \sqrt{g^{FS}}$ hold. We conclude that in the special gauge of Ref.~\cite{TMW} we can describe 3-band kagome model in a geometrical way analogous to the one in the 2-band case. Also, as can be seen from Fig.~\ref{energies_kagome},  in an interval close to
$\phi = \pi/4$ point $M$ has the smallest direct gap among points $K, M$, and $\Gamma$, i.e. points of energy extrema~\cite{f2}. Fig.~\ref{scalarRM} shows that in this interval the scalar curvature, $R^{FS}$, in the special gauge of Ref.~\cite{TMW}, is a decreasing monotonic function of $\phi$ and its values are from the interval $(2, 4)$.
At $\phi = \pi/4$ this value is exactly two.  Therefore, the ``cyclotron" spin that we  infer assuming that the Eq.~(\ref{nonint_eq}) holds in this case is  non-quantized, non-universal (dependent on the parameter of the model - $\phi$) as opposed to the 2-band case. This likely means that the Hall viscosity of the filled band is a non-universal quantity, which can not be expressed in the semi-quantized form (i.e. the result of Ref.~\cite{Hughes2}) of 2-band system in which spin is quantized  and the unit of length is given by the Berry curvature at the point of the effective description.
Note that the gauge of Ref.~\cite{TMW}, in which the geometric description was possible, and in which the
spin for the kagome case was inferred, is  non-periodic in the inverse space and thus physical according to the
Ref. \cite{preprint}.

\section{Conclusions}
\label{Conclusions}
 The Dirac based 2 band CIs like Haldane model are special for having a gauge-invariant Berry
curvature at the expansion (Dirac) point, and that can be described by a local, zero-flux
equation  everywhere, Eq.~(\ref{nonint_eq}) or~(\ref{nonint_eq_sc}). This is a basis for a geometric description
(Section~\ref{The geometric description of the ground state of interacting Dirac based CIs}) of the
interacting problem, near the low-energy expansion point, and allows an introduction of a characteristic length. This
is the length that in the long distance characterizes the size of the particle orbit just as in QHE,
and plays the role of magnetic length in the expression for Hall viscosity (Section~\ref{Discussion} and~\ref{Comparison with other 2 band models}).


In the other model considered, 3 band kagome, we  find a point that has gauge-invariant Berry curvature, supports a geometric description, and satisfies the zero-flux equation with non-quantized and non-universal  ``cyclotron" spin (Section~\ref{3 band kagome model}).

We found that in some regions of phase space of model CIs the geometric description is possible. The ``geometrization" is more probable near QH regions as shown in the two examples, but being in QH region does not guarantee geometrization. Although a gauge-invariant characteristic length may exist, the expansion point may not be the point of the lowest energy gap.  Therefore, with these exemptions and new features, the physics of CIs seems richer than in ordinary QH and provides new mechanisms for the QH phenomena.

Based on the two examples that we analyzed we conjecture that in the Brillouin zone of every band with Chern  number C equal to  $C = 1$ or $C = -1$, we can find a high-symmetry point with the following form of its Bloch vector in its neighborhood,
\begin{equation}
\left[ c_1, c_2, \dots , c_i^x k_x + c_i^y k_y, c_{i+1}, \dots , c_{n+1}\right]^{\mathrm{T}},
\label{conjecture}
\end{equation}
where $ c_i^{x*} c_i^y - c_i^{y*} c_i^x \neq 0$ and $|c_k| = 1/\sqrt{n}$ if $k \neq i$. This is a generalization of the skyrmion expression in Eq.~(\ref{vectorH}) in a two-dimensional $ \mathbf{k}$ space (a plane instead of BZ in the long-distance approximation) to higher band models ant their $|C|=1$ bands. Due to the invariance of the Berry curvature under the diagonal gauge transformations of the expression~(\ref{conjecture}) we may associate with the expansion point the physical characteristic length, $l_D$. This length is connected with the size of the ``skyrmion", $ l_D^2 = -i (c_i^{x*} c_i^y - c_i^{y*} c_i^x)/2$. The characteristic length may characterize the response of the system either in flat (projected) or non-flat bands.


\section*{Acknowledgment}
While we were in the last stage of the preparation of our manuscript, a preprint \cite{preprint} appeared
with some of the claims on the non-invariance of the Berry curvature that are also present in our manuscript.

We would like to thank F.D.M. Haldane, T. Neupert, and N. Regnault for discussions.
This work was supported by the Serbian Ministry of Education and Science under projects No. ON171027, ON171031, and ON171017.

\appendix*
\section{Spin connection and scalar curvature in the space of Bloch vectors}
In this Appendix we derive spin connection and scalar curvature
in the case of 2 and 3 band CI if the equation~(\ref{B-g}) holds. The metric is defined as
\begin{equation}
g_{ij}(k) = w^+_i w_j + w^+_jw_i, \label{gijDef}
\end{equation}
where the two component spinor $w_i$ is defined as $w_i =
(\partial_i u - (u^*\partial_i u) u)/\sqrt{2}= D_i u/\sqrt{2}$
and has components $w_{i\alpha}$. Indices $i,j,\dots$ are space indices and take
values $1,2$, while $\alpha,\beta,\dots$ label components of spinors.

Now we use the relationship between the Berry curvature and the determinant of the
metric tensor
\begin{equation}
B(k) = 2\sqrt{g}.\label{B-g}
\end{equation}
and we notice that
\begin{align}
B(k) &= -\mathrm{i}\left(\partial_x u^+ \partial_y u - \partial_y u^+ \partial_x u\right) \nonumber\\
&= -\mathrm{i}\left(D_x u^+ D_y u - D_y u^+ D_x u\right)\nonumber\\
&= -2\mathrm{i}\epsilon^{ij}w^+_i w_j.\label{B-w}
\end{align}
Combining~(\ref{B-g}) and~(\ref{B-w}) we can define
\begin{equation}
w^j = -\frac{\mathrm{i}}{\sqrt{g}}\epsilon^{jk}w_k,\quad w^{+j} =
\frac{\mathrm{i}}{\sqrt{g}}\epsilon^{jk}w^+_k,
\end{equation}
where $\epsilon^{ij}$ is the totally antisymmetric tensor (Levi-Civita) in two
dimensions.

Then the following relations hold:
\begin{align}
w^{+j}w_j= 1,\quad w^j_\alpha w_{j\alpha} &=
-\frac{\mathrm{i}}{\sqrt{g}}\epsilon^{jk}w_{k\alpha}w_{j\alpha}=0,\nonumber\\
w^{+j}_\alpha w^*_{j\alpha} &=
-\frac{\mathrm{i}}{\sqrt{g}}\epsilon^{jk}w^+_{k\alpha}w^+_{j\alpha}
=0.\label{ww-relations}
\end{align}
In the last two relations there is no sum on $\alpha$, while in the first there
is: $w^{+j}w_j=\sum_\alpha w^{*j}_\alpha w_{j\alpha} = 1$. We also have
\begin{align}
w^+_i w_j &= \frac{1}{2}(g_{ij} + \mathrm{i}\sqrt{g}\epsilon_{ij}), \quad
g^{ij} =\frac{1}{g}\epsilon^{ik}\epsilon^{jl}g_{kl}\nonumber\\
w^{+i} w^j &= \frac{1}{2}(g^{ij} -
\frac{\mathrm{i}}{\sqrt{g}}\epsilon^{ij}),\label{wiwj-relations}\\
w_j &= -\mathrm{i}\sqrt{g}\epsilon_{jk}w^k,\quad w^+_j =
\mathrm{i}\sqrt{g}\epsilon_{jk}w^{+k}.\nonumber
\end{align}

Next we introduce local flat (``Lorentz") indices $a,b, {\dots}=1,2$; they are
related with the metric $\eta_{ab}=\mathrm{diag}(1,1)$. Following Ref.~\cite{YeJe} we construct two dimensional vielbeins, i.e. zweibeins $e_i^{\ a}$ and the
inverse zweibeins $e_a^{\ i}$ from the two component spinors $w_i$:
\begin{align}
e_{i\alpha}^{\ 1} &= \frac{1}{\sqrt{2}}(w_{i\alpha} + w_{i\alpha}^+), \,
e_{i\alpha}^{\ 2} = \frac{\mathrm{i}}{\sqrt{2}}(w_{i\alpha} - w_{i\alpha}^+) \nonumber\\
e_{1\alpha}^{\ i} &= \frac{1}{\sqrt{2}}(w_\alpha^i + w_\alpha^{+i}), \,
e_{2\alpha}^{\ i} = \frac{\mathrm{i}}{\sqrt{2}}(w_\alpha^i - w_\alpha^{+i}).
\label{Zweibeins}
\end{align}
These zweibeins relate space and flat indices and are constructed in such a way
that they are real. The following formulas are valid:
\begin{align}
g_{ij} = \eta_{ab}\sum_\alpha e_{i\alpha}^{\ a}e_{j\alpha}^{\ b},& \quad
\eta_{ab} = g_{ij}\sum_\alpha e_{a\alpha}^{\ i}e_{b\alpha}^{\ j}, \nonumber\\
\sum_\alpha e_{i\alpha}^{\ a}e_{a\alpha}^{\ j} =\delta_i^j,& \quad
\sum_\alpha e_{a\alpha}^{\ i}e_{i\alpha}^{\ b} =\delta_a^b.\label{ge-relations}
\end{align}

Using the metricity condition
\begin{equation}
\nabla_i^{\text{tot}} e_{j\alpha}^{\ a} = \partial_i e_{j\alpha}^{\ a} + \Omega_{i\ \
b}^{\ a}e_{j\alpha}^{\ b} - \Gamma_{ij}^k e_{k\alpha}^{\ a} =0\label{TotKovIzv}
\end{equation}
and the vanishing torsion $\Gamma_{ij}^k = \Gamma_{ji}^k$, we can calculate the
spin connection in terms of zweibeins. From~(\ref{TotKovIzv}) it follows
\begin{equation}
\Omega_{i\ \ b}^{\ a} = \sum_\alpha e_{k\alpha}^{\ a} (\partial_i e_{b\alpha}^{\
k} + \Gamma_{ij}^k e_{b\alpha}^{\ j}) = -\sum_\alpha (\nabla_i e_{j\alpha}^{\
a})e_{b\alpha}^{\ j}, \nonumber
\end{equation}
where $\nabla_i$ is a covariant derivative with respect to the $\Gamma_{ij}^k$
(Christoffel) connection. In equation~(\ref{TotKovIzv}) $\nabla_i^{\text{tot}}$ is the
total covariant derivative of the zweibein, that is with respect to both spin
connection and the $\Gamma_{ij}^k$ connection. We also used $\sum_\alpha
\partial_j(e_{a\alpha}^{\ i})e_{i\alpha}^{\ b} = -\sum_\alpha
e_{a\alpha}^{\ i}(\partial_j e_{i\alpha}^{\ b})$ which follows from
$\sum_\alpha e_{a\alpha}^{\ i}e_{i\alpha}^{\ b} =\delta_a^b$. Since
we are in two dimensions, there are only two independent components of the spin
connection
$\Omega_{i}^{12}$ and $i=1,2$. Inverting relations~(\ref{Zweibeins}), we can
write the spin connection in terms of $w_i$ as
\begin{align}
\Omega_{i}^{\ 12} &= -\frac{1}{\sqrt{g}}\epsilon^{kl}(\nabla_i
w_k^+)w_l\label{SpinKon}\\
&= -\frac{1}{\sqrt{g}}\epsilon^{kl}(\partial_i w_k^+)w_l +
\frac{1}{2\sqrt{g}}\epsilon^{kl}\partial_k g_{il} + \frac{\mathrm{i}}{4}\partial_i(\ln
g).\nonumber
\end{align}

From this spin connection one can calculate the scalar curvature $R$. It is
given by
\begin{align}
R &= R_{ij}^{ab} e_a^{\ i} e_b^{\ j} = 2 R_{ij}^{12} e_1^{\ i} e_2^{\ j} = -\frac{2}{\sqrt{g}}\epsilon^{ij}\partial_i \Omega_j^{12} =
\nonumber\\
& -\frac{2}{\sqrt{g}}\epsilon^{ij}\epsilon^{mn}\left\{ \partial_i
\left(\frac{1}{\sqrt{g}}\right)
\left[-(\partial_j w^+_m)(w_n) + \frac{1}{2}\partial_m g_{jn} \right]\right.
\nonumber\\
& +\left. \frac{1}{\sqrt{g}}\left[-(\partial_j w^+_m)(\partial_i
w_n) + \frac{1}{2}\partial_i\partial_m g_{jn} \right] \right\},\label{SkalKrivina}
\end{align}
where we used that $R_{ij}^{12} = \partial_i \Omega_j^{12} - \partial_j
\Omega_i^{12} = - R_{ij}^{21}$ and $R_{ij}^{11} = R_{ij}^{22} = 0$ because of
antisymmetry.



\begin{thebibliography}{9}
\bibitem{f1} At $\Gamma$ point we find singular behavior.
\bibitem{f2} Other points may have lower direct gap but they are not points of energy extrema.
\end{thebibliography}
\end{document}